\documentclass{article}
\usepackage{spconf,amsmath,graphicx}
\usepackage{hyperref}
\RequirePackage{graphicx}
\RequirePackage{amssymb,amsmath,bm}
\RequirePackage{textcomp}
\RequirePackage{booktabs}
\RequirePackage[textfont=it,tableposition=top]{caption}
\usepackage[square,sort&compress,comma,numbers]{natbib}
\def\L{{\cal L}}
\def\editdist{{\cal E}}

\def\seqW{{\overrightarrow{w}}}
\def\audio{{O}}
\def\mwer{{MWER}}

\title{Audio-attention discriminative language model for ASR rescoring}
\name{Ankur Gandhe, Ariya Rastrow}
\address{\{aggandhe, arastrow\}@amazon.com}
\begin{document}
%\ninept
%
\maketitle
\begin{abstract}
End-to-end approaches for automatic speech recognition (ASR) benefit from directly modeling the probability of the word sequence given the input audio stream in a single neural network. However, compared to conventional ASR systems, these models typically require more data to achieve comparable results. Well-known model adaptation techniques, to account for domain and style adaptation, are not easily applicable to end-to-end systems. Conventional HMM-based systems, on the other hand, have been optimized for various production environments and use cases.  In this work, we propose to combine the benefits of end-to-end approaches with a conventional system using an attention-based \emph{discriminative language model} that learns to rescore the output of a first-pass ASR system. We show that  learning to rescore a list of potential ASR outputs is much simpler than learning to generate the hypothesis. The proposed model results in up to 8\% improvement in word error rate even when the amount of training data is a fraction of data used for training the first-pass system.
% , and propose an attention-based discriminative language model that learns to rescore the output of a first-pass ASR system.
  \end{abstract}
\section{Introduction}
Conventional automatic speech recognition (ASR) systems model the probability of transcription given the input audio as three separate models~\cite{hinton2012deep} - acoustic, pronunciation and language models - each trained with a different object function. In general, the acoustic models are based on a hybrid DNN-HMM structure rand language models are $n$-gram based models. Lately, recurrent neural network (RNN) language models have been shown to outperform $n$-gram based models, due to their ability to capture long-term context~\cite{mikolov2010recurrent, sundermeyer2012lstm}. However, due to their computational complexity, it is often not feasible to use recurrent neural networks in the first pass of a real-time ASR system. Instead, they are often used as second-pass rescorers, as the set of candidate word sequences to rescore is usually small in size ~\cite{liu2014efficient, kumar2017lattice, Raju2019ScalableMC}. \\
% Instead they are often used as second-pass rescorer as the set of candidate word sequences to rescore is usually small in size
% Removed citation for sundermeyer2014lattice
% to language modeling have gained traction in overcoming many limitations of n-Gram models. A Neural Language Model essentially leverages a sequential architecture like Recurrent Neural Network to maximize the likelihood of sequence of words seen in the training data.
More recently, a lot of work has been done in training a single neural network that directly learns how to map input speech signal to word sequences (or graphemes)~\cite{graves2013speech, chan2015listen, chiu2018state}, without the need for training separate models. Although the results are state-of-the-art or close to it on certain tasks, most end to end models are trained on large amounts of transcribed data and its unclear if these models will perform better than conventional systems in a low data regime~\cite{prabhavalkar2017comparison} or in (domain) mismatch scenarios where the underlying distribution for (paired speech and text) training data does not match the test distribution~\cite{coldfusion}.
 Various techniques have been explored to overcome these problems and bring some of the existing advantages of a convention system - fusing external language model trained on abundantly available matched text corpora ~\cite{coldfusion, kannan2018analysis}, contextual biasing~\cite{zhao2019shallow, pundak2018deep} and customizing pronunciations~\cite{phoebe2019} - to the end to end models.\\

In this paper, we combine the simplicity of rescoring conventional ASR models, using neural language models, with the benefits of an attention-based end-to-end model.  More specifically, an RNNLM style model is trained using word-level contextual input, while simultaneously attending to audio. The model is trained  using a minimum word error rate (\mwer) criterion, which learns to rescore the $N$-best hypotheses list from a first pass system. The proposed model can learn, using a \textit{discriminative criterion}, to focus and overcome the errors of the conventional first pass ASR model, having access to both word-level and acoustic-based embeddings. To avoid overfitting and data sparsity issues, we propose to use embeddings generated by a first pass acoustic model. This choice can be relaxed depending on the availability of training data and one can decide to learn the acoustic embeddings (or fine tune the first pass acoustic model embeddings) directly using the {\mwer} criterion. However, using first pass acoustic model embeddings can practically speed up both training and inference. Additionally and since the model is trained using a {\mwer} criterion, it does not have to be confined to a normalized generative form. This can drastically speed up inference, especially when dealing with very large vocabulary ASR systems.

% as a joint distribution of speech signal and corresponding word sequence into 2 parts, Acoustic Model (AM) $P(O/W)$ and Language Model (LM) $P(W)$. The two tasks are largely orthogonal to each other, and thus it makes sense to treat them separately. . \\

\label{sec:intro}

\section{Related Work}
\label{sec:format}
{\mwer} criterion has been previously used in ~\cite{hori2016minimum} to improve the performance of LSTM LMs. They train the model using the {\mwer} criterion but they do not explore using an un-normalized model or a model with attention to audio. ~\cite{guo2019spelling} also explored training a spelling correction model to learn the errors made by a first pass system. However, they did not explore using this model for a conventional system or condition on input audio (e.g. by attending to input audio). Besides, their spelling correction model is a generative model (decoder generates corrected characters in a left-to-right manner). Recently, ~\cite{Chiu2019} proposed to use a listen-attend-and-spell (LAS) model as a second pass rescoring model for output of an RNN-Transducer (RNN-T) first pass. They show significant improvements with LAS as rescoring component, even with a shared encoder. However, the amount of training data used for LAS rescoring was the same as the training data used for first pass model while we explore training the rescoring model with much less amount of data and is important for low-resource languages or domain adaptation with limited amount of in-domain data. The best performance was also achieved when the shared encoder was fine tuned in multi-task fashion (combined loss for both RNN-T and LAS).  Additionally, they do not report results on using such a rescoring scheme with a conventional ASR model.

\section{Audio-attention Recurrent Neural Network Language model}
\label{sec:pagestyle}
A language model estimates the probability of a given sequence of words $\seqW$. N-gram LMs make the assumption that the probability of word $w_i$ depends only on previous $n-1$ words, so that the probability can be written as:
\setlength{\abovedisplayskip}{5pt}
\setlength{\belowdisplayskip}{5pt}
\begin{equation}
\label{eq:lm}
\begin{aligned}
p(\seqW) = \prod_{w_i} p(w_i| w_{i-1}, w_{i-2}, ... w_{i-n+1}) \\
\end{aligned}
\end{equation}
However, recurrent neural networks can model the probability of word given its entire history by applying a compact recurrence architecture on the hidden layer of the model.
\begin{equation}
\begin{aligned}
p(\seqW) = \prod_{w_i} p(w_i| h_{i})
\end{aligned}
\end{equation}
where $h_{i}$ is the hidden representation for the word $w_i$. In a simple recurrent network, the forward step for predicting the probability of word $w_i$ can be described as~\cite{hori2016minimum}:
\begin{equation}
\begin{aligned}
x_i = OneHot(w_{i-1}) \\
h_{i} = \sigma ( W_{ih}^Tx_i +  W_{hh}^T h_{i-1}) \\
p(w_i| w_{<i}) = Softmax(W_{ho}^Th_i)
\end{aligned}
\end{equation}
where $OneHot$ returns a low dimensional representation for a word, $\sigma()$ is an element-wise sigmoid function and the final model probabilities are obtained by applying a softmax function on the activation of the final output. The above set of equation can be easily extended to other types of recurrent models such as LSTMs~\cite{hochreiter1997long}. %, GRUs~\cite{cho2014properties}, etc.

An RNN-based language model is trained by minimizing the cross-entropy loss, which directly maximizes the log-likelihood of training data sequences:
\begin{equation}
\begin{aligned}
\L_{CE} = \sum_w \sum_{i=1}^{L} -log\,p(w_i | w_{i-1},  w_{i-2}, .., w_{0})
\end{aligned}
\end{equation}

\subsection{Minimum word error rate training}
\label{sec:mwe_training}
Using a RNN-based language model during first pass of a real-time ASR system is often infeasible (due to lack of state merging as opposed to $n$-gram models) and most applications use RNNLMs to rescore the output of a first pass ASR system. The most common approach is to generate an n-best list from the first pass and replace or interpolate the language model score of the first pass with a score from the RNNLM. 

As a rescoring model, an RNNLM can be directly trained to rescore the first pass n-best list, as opposed to a maximum likelihood criterion over text-only data. {\mwer} criterion, which estimates an expectation of word errors over an output hypotheses list, has been proposed for both language model training~\cite{hori2016minimum} as well as attention-based models~\cite{chiu2018state}. Let ${\editdist(\seqW_i, \seqW^*)}$  be the edit distance between hypothesis $\seqW_i$ and reference sequence $\seqW^*$, let $\bar{\editdist}$ be the average edit distance for an n-best list, and let $\hat{\editdist}() = \editdist() - \bar{\editdist}$ be the relative edit distance, then {\mwer} loss is defined as:
%~\cite{prabhavalkar2018minimum}.
 \begin{equation}
\begin{aligned}
\L_{werr} = \sum_{\seqW_i \in \textbf{w}} p_{\theta}(\seqW_i|O) \hat{\editdist}(\seqW_i, \seqW^{*})
\end{aligned}
\end{equation}
% % \expectation_{p_{\theta}}[\hat{\editdist}(\textbf{w}, w^{*})] \approx \\
% the summation is over all paths $\seqW_n$ in the first-pass n-best list, $\hat{\editdist}$ = $\editdist() - \bar{\editdist}$  is the edit distance between two word sequences, $w^*$ is the reference word sequence, and
where $p_{\theta}(\seqW_i|O)$ is the normalized posterior probability (over n-best list) of the hypothesis given the input acoustics, computed considering both the acoustic model and the language model probabilities:
\begin{equation}
\label{eq:posterior_prob}
\begin{aligned}
p_{\theta}(\seqW_i|O) = \frac{exp(g_{i})}{\sum_{j=1}^{N} exp({g_{j}})} \\
\end{aligned}
\end{equation}
where
\begin{equation}
\begin{aligned}
g_{i} = \alpha log\,p_{lm}(\seqW_{i}) + log\,p_{am} (O|\seqW_{i})
\end{aligned}
\end{equation}
and $p_{am} (O|\seqW_{i})$ is the sequence probability computed by the acoustic model, $p_{lm}(\seqW_{i})$ is the sequence probability computed by the rescoring RNNLM. $N$ is the size of n-best list. According to equation~\ref{eq:posterior_prob},  the posterior probabilities are normalized over the n-best list. Therefore, it is not necessary for the RNNLM to produce normalized probabilities for each word of the sequence. In this paper, we explore both a normalized and unnormalized output for the RNNLM. \\
To stabilize training and achieve better convergence, we interpolate the word error rate losses with cross-entropy loss, weighted by a hyper-parameter $\lambda$~\cite{chiu2018state}: 
%~\cite{prabhavalkar2018minimum}:
\begin{equation}
\begin{aligned}
\L_{total} = \L_{werr}  + \lambda \L_{CE}
\end{aligned}
\end{equation}

\subsection{Attention to audio}
As described in Section~\ref{sec:intro}, unlike conventional ASR systems with a separate acoustic model and language model, end-to-end systems benefit from directly optimizing the $p(w|\audio)$. Similarly, the input to the standard RNNLM can be modified so that it also models $p(w|\audio)$:
\begin{equation}
\label{eq:attention_to_audio}
\begin{aligned}
P(\seqW | \audio) = \prod_{|\seqW|} p(w_i| h_{i}, \audio) \\
\end{aligned}
\end{equation}
There is no direct alignment between the words and segments of input audio. The dependence in equation~\ref{eq:attention_to_audio} is learnt by using an LSTM with attention to input audio frames. At every step, the probability of next word is computed conditioned on all the previous words as well an attention context over the entire audio sequence:
\begin{equation}
\begin{aligned}
c_i = AttentionContext(h_{i-1}, \audio) \\
h_i = RNN(h_{i-1}, w_{i-1}, c_{i-1}) \\
p(w_i|w_{<i}, O) = Softmax(W_{ho}^T(h_i, c_i))
\end{aligned}
\end{equation}
The attention context over (frame-by-frame) encoded audio is learnt similar to the the way described in ~\cite{chan2015listen}. However, unlike end-to-end systems, we make two distinctions:
\begin{enumerate}
\item Learning a separate encoder over raw audio features is not a requirement for our model. Instead, we directly take the intermediate layers of the first pass acoustic model as input to our model. This reduces the amount of training data needed to learn a good model. In section~\ref{sec:expts} we compare using different encoder types on top of the output of first-pass acoustic model.
%We don't learn a separate encoder from raw audio features. Instead, we attention to the intermediate layers of the first pass acoustic model. This reduces the amount of training data needed to learn a good model. In section~\ref{sec:expts} we compare using different encoder types on top of the first-pass acoustic model
\item The model is not trained to generate the correct transcription but instead trained only to minimize the word error rate of an n-best list generated by a first pass system. This is is an important distinction as the \emph{discriminative} flavor of our model enables it to focus on correcting first-pass errors, as opposed to learn to generate words.
\end{enumerate}

\begin{figure}
\includegraphics[width=\linewidth]{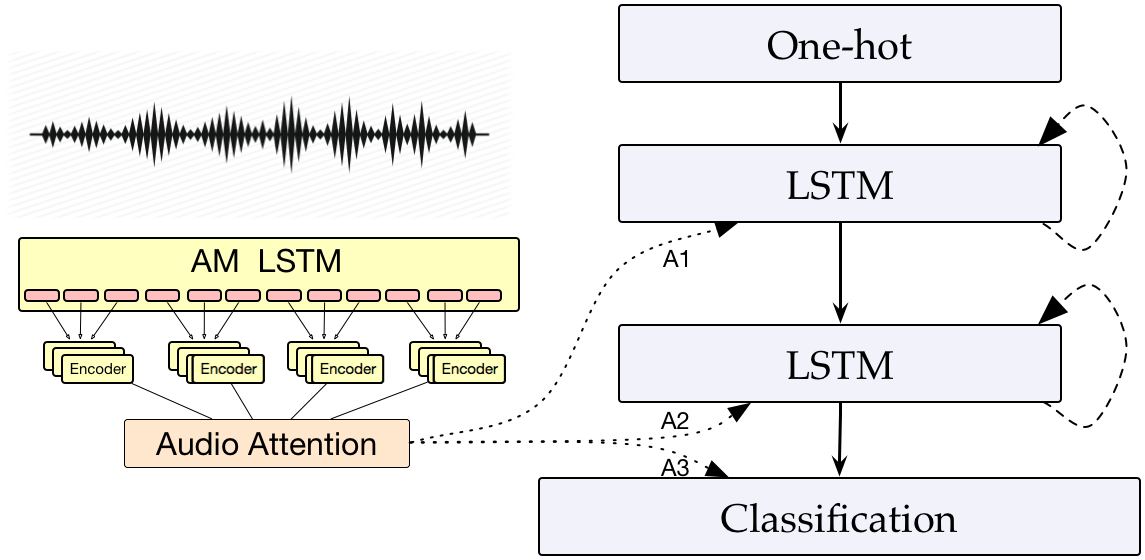}
\caption{LSTM language model with attention to audio embedding from the first pass}
\label{fig:audio_attention_NLM}
\end{figure}

Figure~\ref{fig:audio_attention_NLM} shows the proposed architecture with different ways of adding attention to the LM LSTMs; AM LSTM is the first pass model used for generating the audio embedding, CNN as an optional encoder that can be learnt during model training. 

\section{Experiments}
\label{sec:expts}
In all of the experiments in this paper, we build an ASR system that targets a spoken dialog system used in digital assistants such as Amazon Alexa, Google assistant, Siri, etc. The acoustic model is a low-frame-rate model with 2-layer frquency LSTM~\cite{li2017acoustic}
followed by a 5-layer time LSTM trained on cross-entropy loss, followed by sMBR loss~\cite{vesely2013sequence}.
% ~\cite{li2015lstm} ~\cite{lfr2016} -> replaced with ~\cite{li2017acoustic}

All recurrent neural models (NLM) models have a word embedding size of 512, and comprise two LSTM \cite{hochreiter1997long} layers, each with 512 hidden layers. During training, the models are trained on either cross-entropy loss (NLM-XENT) or {\mwer}  loss (NLM-MWE) as described in section~\ref{sec:mwe_training}, with a maximum n-best list size of 64. During inference, the NLMs are used to rescoring 10-best hypotheses generated from the first-pass decoding. For NLM models with attention to audio (NLM-AUDIO), we attend to the activation of the last LSTM layer (dimension of 768) of the first pass acoustic model and learn to projection it to a smaller dimension of 200. For all experiments, a CNN-based encoder was used to learn the acoustic embeddings (details in section~\ref{sec:expts_encoder}).

The first-pass LM is trained on a variety of out-of and in-domain corpora, written text data and transcribed speech data from real user-agent interactions\footnote{The user data is anonymized and only a subset of data used for our production system} respectively.  The written text corpora contain over 50 billion words in total.
A Kneser-Ney (KN) \cite{kneser1995improved} smoothed n-gram language model is estimated from each corpus with a vocabulary of 400k, and the final first-pass LM is a linear interpolation of these component LMs. The interpolation weights are estimated by minimizing the perplexity on a development set.  The transcribed speech data, used to train the rescoring NLM-XENT comprises of approximately 300 million words of text. A smaller dataset of 5 million tokens is used for training the models that require audio data (NLM-MWE and NLM-AUDIO) in addition to the transcription. From this corpus, we extract the vocabulary of 66k most frequent words. All NLM models use this vocabulary and out of vocabulary tokens are mapped to \verb|<unk>|. 
% Note that the first-pass ASR system has a larger vocabulary of 400k. 
% In rescoring experiments (but not for perplexity computation), we scale the probability of \verb|<unk>| token by a factor of $10^{-6}$, i.e., we assume a uniform distribution over the "missing" vocabulary.
%The amount of training data used for different LMs is shown in table~\ref{table:data}

%\begin{table}[th]
%  \caption{Language model training data for different models}
%  \label{table:data}
%  \centering
%  \footnotesize
%    \begin{tabular}{l | c | c | c}
%    \toprule
%   & \textbf{First-pass LM} & \textbf{NLM} & \textbf{pre-trained NLM}\\
%        \midrule
%    \# of tokens & 50 Billion & 5 Million & 150 Million   \\ % 16.0
%    \bottomrule
%  \end{tabular}
%\end{table}

We evaluate each model on a 40k utterance test data set and measure the  word error rate reduction (WERR) relative to the baseline model.
Table ~\ref{table:wer-numbers} shows the result for different rescoring models both in terms of perplexity\footnote{Perplexity is only reported for normalized models.} and word error rate improvements (compared against the baseline $4$-gram LM).
\begin{table}[th]
  \caption{Perplexity(PPL) and Relative Word Error Rate Reduction (WERR) for different rescoring models}
  \label{table:wer-numbers}
  \centering
  \footnotesize
    \begin{tabular}{l | c | c }
    \toprule
    % \multicolumn{1}{c|}{\textbf{Model}} & \multicolumn{1}{c|}{\textbf{PPL}} & \multicolumn{1}{c|}{\textbf{WERR}} &\multicolumn{2}{c}{\textbf{Rescoring latency}} \\
    % & & & P50 & P90 \\
       \textbf{Model} & \textbf{PPL} & \textbf{WERR} \\
        \midrule
    KN-4g & 37.2 &   \\ % & - & - \\ % 17.1
    % NLM-XENT & 58.7 & 1.1\%  \\ % & -  & - \\ % 16.9
    NLM-XENT & 34.7 & 1.9\%  \\ % & -  & - \\ % 16.9
    NLM-MWE-NORM& 90.7  & 2.6\% \\ %& 767ms & 3396ms\\ % 16.3
    NLM-MWE& -  & 3.6\% \\  % & 767ms & 3396ms\\ % 16.3
    NLM-MWE-AUDIO-NORM & 18.5 & 7.0\% \\ %& 10ms & 3396ms\\ % 15.8
    NLM-MWE-AUDIO & - & 7.3\% \\ % & 10ms & 3396ms\\ % 15.8
    10-best oracle & - & 28\% \\ %& - & - \\
    \bottomrule
  \end{tabular}

\end{table}
% Due to the nature of the task, where the average number of words in each utterance is 5, 
We see very small improvement when using a cross-entropy trained NLM. However, with {\mwer} rate training with un-normalized output, even on much smaller amount of data, the improvement is 3.6\%; the perplexity of the normalized is much higher  because its not optimized for maximum likelihood. 
Further, the model with attention to acoustic embeddings, from the first-pass AM results in about 7.3\% improvement. The perplexity of the NLM model with attention to audio is significantly low because its able to attend to the audio frames corresponding to the word being predicted. We also see that the un-normalized models perform slightly better than the normalized model because they have less constraints on the output scores.

\subsection{Using pre-trained language model}
Minimum word error rate training requires audio data to generate the n-best list required for training. However, LMs trained with cross-entropy can be trained on much larger amounts of text-only data. Table ~\ref{table:pre-trained-lm} shows the result of fine-tuning a pre-trained NLM-XENT (trained on 300 million tokens) with {\mwer} criterion.
%on large amounts of text-only data and then fine-tuning the model with minimum word error rate. The pre-trained model was trained on 300 million token corpus of text-only transcriptions. 
In case of audio-attention model, it is not possible to pre-train all the layers of model, since it needs an extra context input. Here, we pre-train the two LSTM layers of the model and apply audio context only to the final affine. We see only a minor improvement over the randomly initialized model in all cases. This might be due to the amount of audio training data we have. \\
\begin{table}[th]
  \caption{Perplexity(PPL) and Relative Word Error Rate Reduction (WERR) for different attention location}
  \label{table:pre-trained-lm}
  \centering
  \footnotesize
    \begin{tabular}{l | c | c | c | c }
    \toprule
   %  \multicolumn{1}{c|}{\textbf{Model}} & \multicolumn{1}{c|}{\textbf{PPL}} & \multicolumn{1}{c|}{\textbf{WERR}} &\multicolumn{2}{c}{\textbf{Rescoring latency}} \\
 & \multicolumn{2}{c|}{No-pretrain} & \multicolumn{2}{c}{Pretrain} \\
%             & PPL            & WER            & PPL           & WER          \\
   \textbf{Model} & \textbf{PPL} & \textbf{WERR} & \textbf{PPL} & \textbf{WERR} \\
        \midrule
    NLM-XENT & 34.7 & 1.9 \% & 33.8 & 1.9\%  \\ % 16.0
    NLM-MWE & - & 3.6\% & -  & 4.5\% \\ % 16.0
    NLM-MWE-AUDIO & - & 7.3\% & - & 7.7\% \\ % 15.82
    \bottomrule
  \end{tabular}
\end{table}
% Figure ~\ref{fig:learning_curve} shows the improvement using MWER training for language model with and without audio attention. We see that even with a small amount of training data, the method helps in reducing the word error rate.

\subsection{Adding audio attention to LM}
\label{sec:expts_attention_location}
In a standard end-to-end systems, the attention to encoder states is applied at every layer. However, in a language model, the model has to learn both the dependency to acoustics as well as the temporal dependency in the word sequence. Table  ~\ref{table:attention-location} shows results of different variations of attention location, as shown in figure~\ref{fig:audio_attention_NLM}. We see that adding audio context to the last LSTM-LM layer leads to some degradation in performance. However, using audio context at final affine layer or at the first LSTM layer leads to similar results.

\begin{table}[th]
  \caption{Model size and Relative Word Error Rate Reduction (WERR) for different attention location}
  \label{table:attention-location}
  \centering
  \footnotesize
    \begin{tabular}{l | c | c }
    \toprule
   %  \multicolumn{1}{c|}{\textbf{Model}} & \multicolumn{1}{c|}{\textbf{PPL}} & \multicolumn{1}{c|}{\textbf{WERR}} &\multicolumn{2}{c}{\textbf{Rescoring latency}} \\
   \textbf{Attention location} & \textbf{\# params (MM)} & \textbf{WERR} \\
        \midrule
    Final Affine (A3)  & 69 & 7.3\% \\ % 15.85
    First LSTM layer (A1) & 71 & 6.6\%  \\ % 15.9
    Last LSTM layer (A2) & 71 & 6.1\% \\ % 16.1
    First LSTM + Final Affine  & 87 & 7.4\% \\ % 15.8
    \bottomrule
  \end{tabular}
\end{table}

\subsection{Learning context encoder for audio embeddings}
\label{sec:expts_encoder}
In addition to attending to hidden layer outputs of the first pass LSTM-AM, we also experimented with applying a learnable context encoders on top of the LSTM AM outputs. Following various papers, we experimented with three different encoder types: a) 2-layer pyramidal LSTM (PyLSTM)~\cite{chan2015listen} b) single layer time-delay Neural Network (TDNN)~\cite{peddinti2015time} with context $\{-1, 2\}$ and c) convolutional neural network (CNN) with filter size 3x3 and non-overlapping max pooling of size 3~\cite{sainath2015convolutional}. \\
Table ~\ref{table:context-encoder} shows the result of applying these different context encoders. We see that using Pyramidal LSTM and TDNNs for encoding the input improves the performance of the model.  However, both TDNN and Pyramidal LSTM have a significant increase in number of parameters as well as training and inference time. CNN encoders are able to improve the performance over having no encoder. For all our other experiments, we chose to go with CNN encoder because of its fast training and inference speed.

\begin{table}[th]
  \caption{Model size and Relative Word Error Rate Reduction (WERR) for different attention location}
  \label{table:context-encoder}
  \centering
  \footnotesize
    \begin{tabular}{l | c | c }
    \toprule
   \textbf{Encoder type} & \textbf{\# params (MM)} & \textbf{WERR} \\
        \midrule
    No encoder & 68 & 7\%  \\ % 16.0
    PyLSTM & 84  & 7.9\% \\ % 16.0
    TDNN  & 83 & 8.0\% \\ % 15.82
    CNN & 69 & 7.3\% \\ % 15.8
    \bottomrule
  \end{tabular}
\end{table}

%\subsection{Affect of subword units}
%\label{sec:expts_subwords}
%Similar to experiments done in ~\cite{chan2015listen}, we also experimented with different sub-word language models for second-pass rescoring. We used byte-pair encoding for learning the subwords. Note that the first pass ASR model is still a word level model. Table ~\ref{table:subwords} shows the word error rate results from using different sub-word models
%
%\begin{table}[th]
%  \caption{Perplexity(PPL) and Relative Word Error Rate Reduction (WERR) for different attention location}
%  \label{table:subwords}
%  \centering
%  \footnotesize
%    \begin{tabular}{l | c | c }
%    \toprule
%   %  \multicolumn{1}{c|}{\textbf{Model}} & \multicolumn{1}{c|}{\textbf{PPL}} & \multicolumn{1}{c|}{\textbf{WERR}} &\multicolumn{2}{c}{\textbf{Rescoring latency}} \\
%   \textbf{Sub-word size} & \textbf{PPL} & \textbf{WERR} \\
%        \midrule
%    Word-level & - & 7.6\%  \\ % 16.0
%    4k  & -  & 7.4\% \\ % 16.0
%    8k  & - & 7.8\% \\ % 15.82
%    16k & - & 7.6\% \\ % 15.8
%    \bottomrule
%  \end{tabular}
%\end{table}

\section{Conclusion}
\label{sec:conclusion}
In this paper, we proposed to use an attention-based language model for second-pass rescoring of n-best lists generated by a conventional ASR system. We show that by training the model with minimum word error rate (MWER) criteria, we can get upto 4.5\% word error rate improvement over the the first pass system. Further, we show that the attention-based model can improve the word error rate even when the amount of training data is less and no task-specific audio encoder is learnt. We also show that pre-training the word embeddings and LSTM layers of the model can improve the performance of the model to 8\% WERR. Future work will compare other end-to-end systems with this approach as well as extending this work to very low-data regime such as few-shot learning.

\bibliographystyle{IEEEbib}
\bibliography{refs}

\end{document}